\documentclass[sigconf]{acmart}

\AtBeginDocument{%
  \providecommand\BibTeX{{%
    \normalfont B\kern-0.5em{\scshape i\kern-0.25em b}\kern-0.8em\TeX}}}

\copyrightyear{2022}
\acmYear{2022}
\setcopyright{rightsretained}
\acmConference[CHI '22]{CHI Conference on Human Factors in Computing Systems}{April 29-May 5, 2022}{New Orleans, LA, USA}
\acmBooktitle{CHI Conference on Human Factors in Computing Systems (CHI '22), April 29-May 5, 2022, New Orleans, LA, USA}
\acmDOI{10.1145/3491102.3501903}
\acmISBN{978-1-4503-9157-3/22/04}

\usepackage{soul}
\usepackage{float}
\usepackage{multirow}
\usepackage{epstopdf}
\usepackage{hyperref}
\usepackage{listings}
\usepackage{fancybox}
\usepackage{graphicx}
\usepackage{subcaption}
\usepackage{ragged2e}
\usepackage{enumitem}

\usepackage{amsmath,bm}

\usepackage[framemethod=TikZ]{mdframed}
\usepackage{tikz}
\usepackage{pgfplots}
\usetikzlibrary{pgfplots.statistics,calc}
\usepackage{color}
\usepackage{xcolor}
\usepackage{array}
\usepackage{amsmath}
\usepackage{centernot}
\usepackage{xspace}
\usepackage{url}
\usepackage{verbatim}
\usepackage{wrapfig}
\usepackage{tabularx}
\usepackage{bbding}
\usepackage{pifont}
\usepackage{wasysym}
\usepackage{amssymb}

\newcommand{\tool}{\texttt{NaviDroid}}

\makeatletter  
\newif\if@restonecol  
\makeatother  
\usepackage[linesnumbered,ruled,vlined]{algorithm2e}
\usepackage{algpseudocode}  
\usepackage{amsmath}  

\begin{document}

\title{Guided Bug Crush: Assist Manual GUI Testing of Android Apps via Hint Moves}

\author{Zhe Liu}
\authornote{Also With University of Chinese Academy of Sciences, Beijing, China; \\ Laboratory for Internet Software Technologies, Beijing, China; \\ State Key Laboratory of Computer Sciences, Beijing, China;\\ Science \& Technology on Integrated Information System Laboratory, Beijing, China}
\affiliation{%
\institution{Institute of Software \\ Chinese Academy of Sciences}
\city{Beijing}
\country{China}
}
\email{liuzhe2020@iscas.ac.cn}

\author{Chunyang Chen}
\affiliation{%
  \institution{Monash University}
  \city{Melbourne}
  \country{Australia}
}
\email{Chunyang.chen@monash.edu}

\author{Junjie Wang}
\authornotemark[1]
\authornote{Corresponding author}
\affiliation{%
\institution{Institute of Software \\ Chinese Academy of Sciences}
\city{Beijing}
\country{China}
}
\email{junjie@iscas.ac.cn}

\author{Yuekai Huang}
\authornotemark[1]
\affiliation{%
\institution{Institute of Software \\ Chinese Academy of Sciences}
\city{Beijing}
\country{China}
}
\email{yuekai@iscas.ac.cn}

\author{Jun Hu}
\authornotemark[1]
\affiliation{%
\institution{Institute of Software \\ Chinese Academy of Sciences}
\city{Beijing}
\country{China}
}
\email{hujun@iscas.ac.cn}

\author{Qing Wang}
\authornotemark[1]
\authornotemark[2]
\affiliation{%
\institution{Institute of Software Chinese Academy of Sciences}
\city{Beijing}
\country{China}
}
\email{wq@iscas.ac.cn}

\begin{abstract}
Mobile apps are indispensable for people's daily life.
Complementing with automated testing, manual testing is the last line of defence for app quality. 
However, the repeated actions and easily missing of functionalities make manual testing time-consuming and inefficient.
Inspired by the game candy crush with flashy candies as hint moves for players, we propose an approach named {\tool} for navigating testers via highlighted next operations for more effective and efficient testing.
Within {\tool}, we construct an enriched state transition graph with the triggering actions as the edges for two involved states. Based on it, we utilize the dynamic programming algorithm to plan the exploration path, and augment the GUI with visualized hints for testers to quickly explore untested activities and avoid duplicate explorations.
The automated experiments demonstrate the high coverage and efficient path planning of {\tool} and a user study further confirms its usefulness.
It can help us develop more robust software that works in more mission-critical settings, not only by performing more thorough testing with the same effort that has been put in before, but also by integrating these techniques into different parts of development pipeline.

\end{abstract}


\maketitle

\section{Introduction}
\label{sec_introduction}
Due to the portability and convenience of mobile phones, they are more and more popular in the current world.
Given 3 million available Android applications (apps)~\cite{Number2020Google} for different tasks such as reading, shopping, banking and chatting~\cite{Mobile2016first}, mobile phones and apps now have become indispensable for our daily life in accessing the world~\cite{chen2018ui}. 
The importance of mobile apps makes it vital for the development team to carry out a thorough testing for ensuring the quality of mobile apps.
The quality of the mobile application also decides its success among many similar apps in the market.

However, it is challenging to guarantee the mobile application quality, especially considering that mobile applications are event-centric programs with rich graphical user interfaces (GUIs)~\cite{wu2019analyses}, and interact with complex environments (e.g., users, devices, and other apps). 
To ensure the application quality, there are mainly two kinds of GUI testing i.e., automated GUI testing and manual GUI testing.
There have been many automated GUI testing studies for mobile applications including model based~\cite{mirzaei2016reducing,yang2018static,yang2013grey}, probability based~\cite{machiry2013dynodroid,zeng2016automated,mao2016sapienz} and deep learning based~\cite{li2019humanoid,pan2020reinforcement} approaches. 
Most of them are dynamically exploring mobile apps by executing different actions such as scroll and click with random input, based on the analysis of the code structure of the current page.
Albeit its convenience and scalability, it also has the following challenges.
For example, automated testing may not have a high activity coverage, especially the deeper UI pages and functions are difficult to cover~\cite{mao2016sapienz,machiry2013dynodroid,van2010automating}. Complex operations are difficult to implement, especially for those functionalities which can only be reached by complicated inputs or a long sequence of actions~\cite{cai2020fastbot,wiklund2017impediments,su2017guided}. Furthermore, the usability and accessibility bugs (e.g., color schema, font size, interaction)~\cite{chen2021accessible,yang2021uis,yang2021don} are difficult to reveal by automated GUI testing~\cite{alshayban2020accessibility,fazzini2017automated,ciupa2008finding,DBLP:conf/chi/SalehnamadiAL0B21,DBLP:conf/chi/PetrieK07,DBLP:conf/chi/PowerFPS12}.

\begin{figure*}[htb]
\centering
\includegraphics[width=16.6cm]{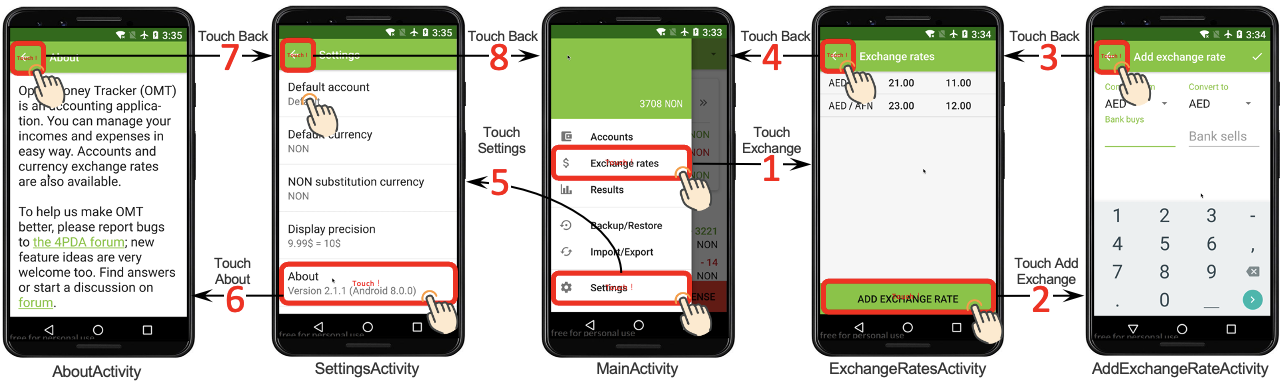}
\vspace{-0.1in}
\caption{Example of the {\tool} usage scenario.}
\label{fig:NaviDroid_example}
\vspace{-0.1in}
\end{figure*}

Therefore, in addition to automated GUI testing, companies also adopt manual testing as the last line of defence ~\cite{linares2017developers,ISTQB2018,leitner2007reconciling,engstrom2010systematic}.
Research also showed that due to the usability and learning curve of automated tools, manual testing is preferred by many software developers~\cite{linares2017developers,DBLP:conf/sigsoft/HaasEJPA21,DBLP:conf/chi/Garcia20a}.
Within the manual testing, multiple testers mimic users' behaviors to explore different functionalities in the apps aiming to find more bugs~\cite{itkonen2009testers,liu2020owl,ISTQB2018,DBLP:conf/sigsoft/SuLC0W21}.
Some researchers optimize manual testing according to test cases and prioritization, but these techniques usually rely on specific data~\cite{DBLP:conf/icst/HemmatiFM15,DBLP:conf/icst/HemmatiS18,DBLP:conf/icmla/LachmannSNSS16}.
To harness crowdsourcing's diversity, crowdsourced testing~\cite{itkonen2014test,rafi2012benefits,wang2019iSENSE} recently emerges in software testing which exploits the benefits, effectiveness, and efficiency of crowdsourcing and the cloud platform, to replace conventional manual testing which limits the fixed number of testers.
Compared with automated GUI testing, human testers are able to discover more diverse and complicated bugs, especially those related to user experience.
But there are also two problems with the manual GUI testing.
First, it is time-consuming which requires a large number of testers to manually explore each screen of the app, and testers may execute repeated actions during the exploration~\cite{DBLP:conf/sigsoft/HaasEJPA21,DBLP:conf/icst/ElodieBALU20,DBLP:conf/icse/EderHJVP14}. 
Second, the performance of manual testing is unstable as it highly depends on the testers' capability and experience, and testers may miss some minor functionalities especially for those unfamiliar apps~\cite{DBLP:conf/icst/HemmatiS18,8990261,DBLP:conf/sigsoft/HaasEJPA21}.
Therefore, the target users of this paper are the testers who are not familiar with the application they are testing.

There are always pros and cons of automated GUI testing and manual GUI testing, and separating them into two unrelated processes may further deepen their cons~\cite{linares2017developers,itkonen2009testers}. 
To leverage the pros of both testing techniques, we propose a new hybrid method to assist manual GUI testing based on the insights from the automated GUI testing.
Inspired by the automated GUI testing, we first distill the prior knowledge of one app including all states and state relationships.
We then implement that prior knowledge into a tool {\tool}.
During manual GUI testing, our {\tool} will trace testers' testing steps and help navigate or remind testers with unexplored pages by explicit visual annotations (e.g., red bounding box) in the run-time page as seen in Fig \ref{fig:NaviDroid_example}.
Our {\tool} can help human testers avoid missing some functionalities or making repeated exploration steps.

Within our approach, there are mainly three components including distilling prior knowledge via the program analysis, planning the exploration path, and providing visual-based path guidance for testing the applications.
\textbf{\textit{First}}, $STG_{action}$ is constructed, which is a state transition graph with each edge annotated with the trigger action between two states (e.g., clicking the ``login'' button in the state \textit{login} to the next state \textit{landingPage}). 
We combine the static program analysis and dynamic random exploration to ensure the states and trigger actions are accurately captured.
We also design a context-aware state merging method to merge the near-duplicate states, by considering both the current state and the adjacent states.
\textbf{\textit{Second}}, based on the extracted graph, we utilize dynamic programming algorithm to plan the exploration path to efficiently cover all the states with few repeated exploration steps.
\textbf{\textit{Third}}, following the planned path, we augment the run-time GUI with visual hint moves to provide real-time guidance in mobile GUI testing.
That process is similar to the flashing candies (hint/suggested moves) when a player hesitates to make a move in playing Candy Crush (a popular free-to-play match-three puzzle video game)~\cite{Candy2020}.
According to our observation, that suggested move is particularly useful when the trigger components are small or poorly designed/developed. 

As the {\tool} consists of STG extraction algorithm, dynamic programming algorithm and visual guidance. Therefore, we first evaluate the algorithm performance of the {\tool} through an automated method. We evaluated the {\tool} on 85 open-source apps from F-Droid (the largest repository of open-source Android Apps). 
Results show that {\tool} can achieve 74\% median activity coverage and 81\% median state coverage with the extracted $STG_{action}$, outperforming five commonly-used and state-of-the-art baselines. 
It also saves 20\% to 42\% exploration steps compared with the three commonly-used baselines. 
We further carry out a user study to evaluate its usefulness in assisting manual GUI testing, with 20 apps from F-droid. 
Results show that, the participants with {\tool} cover 62\% more states and 61\% more activities, detect 146\% more bugs within 33\% less time, compared with those without using {\tool}.
This confirms the usefulness of {\tool} in avoiding missing functionalities and making repeated exploration steps, and helping detecting bugs during manual testing.
The demo video link is \url{https://youtu.be/7kR9-9-gPQ0}.
The contributions of this paper are as follows:

\begin{itemize}
\item The first Android manual testing assistant {\tool}\footnote{We release the source code, experiment detail, and demo videos of our {\tool} in https://github.com/20200829/Navidroid. The demo video link is \url{https://youtu.be/7kR9-9-gPQ0} \label{github}}, to the best of our knowledge, by providing visual-based exploration path guidance based on the states and trigger actions, which can avoid missing functionalities or making repeated exploration steps.

\item $STG_{action}$ extraction method, which combines static analysis and dynamic exploration to acquire the states and related trigger actions, and incorporates context-aware state merging method for near-duplicate state merging.  

\item The DP-based (dynamic programming) path planning method, which guides the path exploration in covering all the states with few repeated exploration steps.

\item Effectiveness evaluation on real-world mobile apps with promising results, and a user study demonstrating the usefulness of {\tool} in assisting manual GUI testing and finding practical bugs.

\end{itemize}

\section{Related Work}
\label{sec_related}
With the rapid growth of the number of apps, their functions are also increasing. 
This leads to the researchers being more and more concerned about bugs from Android apps~\cite{wu2019analyses}. 
GUIs are the primary UI in the vast majority of today's commodity software~\cite{liu2020owl,mazza2017reducing,chen2021should,feng2021auto,zhao2021guigan}.
However, creating GUI tests to cover a large number of Android UI pages is a challenging task.

\textbf{Automated GUI Testing.}
To save human efforts in manual testing, many researchers explored the automatic generation approaches of large-scale test scripts for application testing~\cite{xie2007designing}.
There are a number of linting tools~\cite{lint,stylelint,yang2018static} based on the static program analysis to mark programming errors, bugs, style errors, and suspicious constructs to ensure that the application works properly.
Due to the complexity of mobile apps, random GUI testing techniques based on dynamic exploration are proposed~\cite{mao2016sapienz,cai2020fastbot}, which aims to cover more components or activities. 
These dynamic GUI testing tools~\cite{borges2018droidmate,li2019humanoid, su2017guided,zhao2020seenomaly} can simulate human operation (e.g., click, long press, slide, etc.) to test the application.

\begin{figure*}[htb]
\centering
\includegraphics[width=15.5cm]{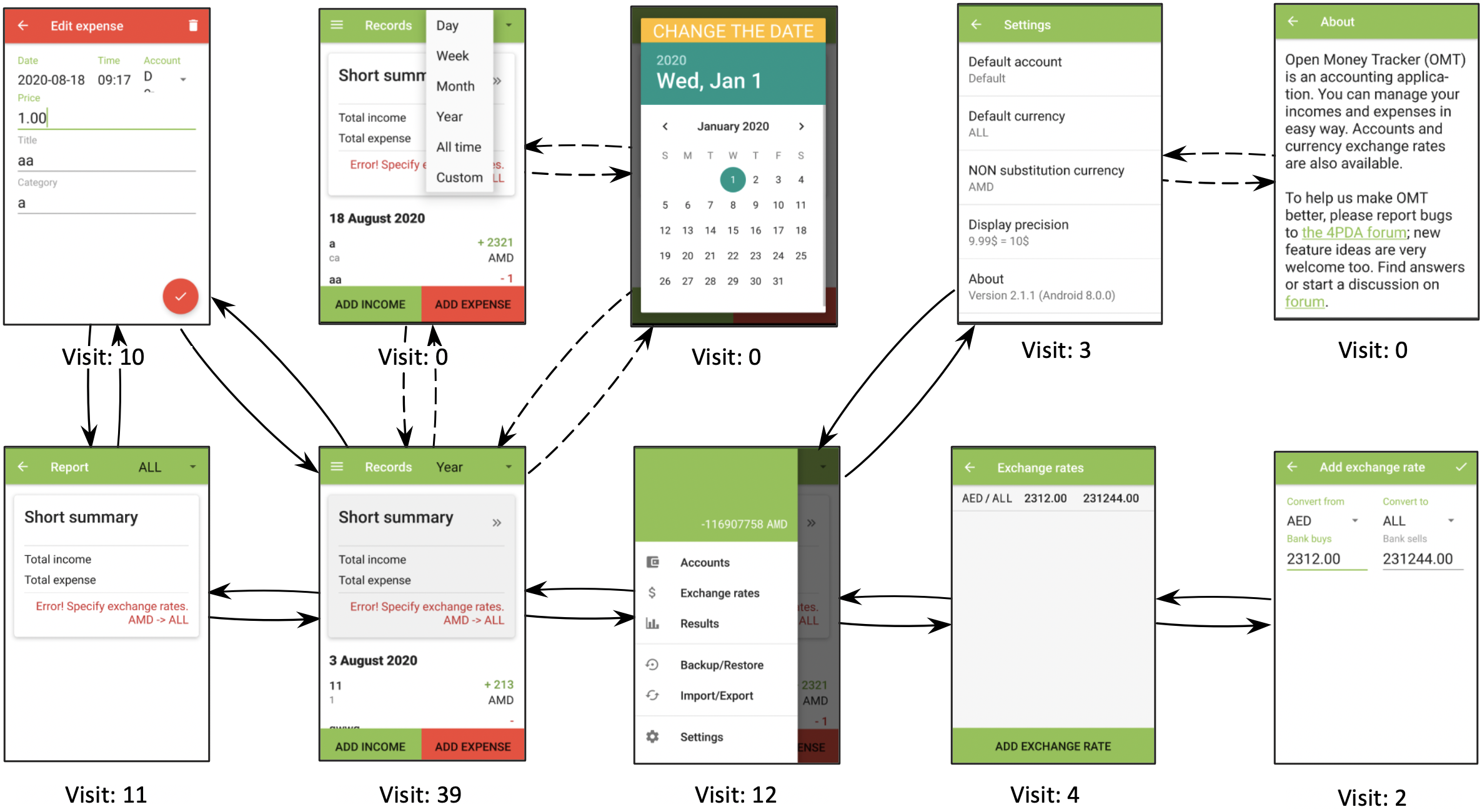}
\caption{Example of one tester's exploration graph. The solid line represent the explored path while dotted line is the unexplored path. Number below each page is the visit time of the tester.}
\vspace{-0.1in}
\label{fig:example_motivation}
\vspace{-0.05in}
\end{figure*}

Although automated GUI testing can quickly spot functionality bugs of mobile apps, it has limitations in detecting the bugs related to complicated sequential operations or about app usability and accessibility~\cite{fazzini2017automated,alshayban2020accessibility,DBLP:conf/chi/SalehnamadiAL0B21,ciupa2008finding,DBLP:conf/chi/PetrieK07,DBLP:conf/chi/PowerFPS12}, which requires manual testing. 
And the automated GUI testing may not have a high activity coverage, such as the functions and interface that are easy to be ignored are difficult to cover~\cite{mao2016sapienz,machiry2013dynodroid,van2010automating}. Complex operations are difficult to implement, especially for those functionalities which can only be reached by complicated inputs or a long sequence of actions~\cite{cai2020fastbot,wiklund2017impediments,su2017guided}.
Empirical studies show that companies still rely on manual testing~\cite{itkonen2009testers,mao2017crowd,linares2017developers,ISTQB2018,leitner2007reconciling,engstrom2010systematic,DBLP:conf/sigsoft/HaasEJPA21,DBLP:conf/chi/Garcia20a}. The test execution is not a simple mechanical task, but a creative and experience-based process, and manual testing will never be replaced by automatic testing.

\textbf{Manual GUI Testing.}
Manual testing can be defined as a process in which software testers manually verify the correctness of software functionalities according to the requirements provided by customers~\cite{linares2017continuous, linares2017developers,danilov2016software}. 
Many studies explored the factors influencing testers' performance such as training~\cite{micallef2016exploratory,whittaker2009exploratory}, recruitment process~\cite{cui2017multi,xie2017cocoon}, and their experience~\cite{itkonen2009testers}.
However, all these studies are based on pure manual exploration which highly depends on the expertise and experience of testers.
Some studies also optimize manual testing according to task priority~\cite{DBLP:conf/icst/HemmatiFM15,DBLP:conf/icst/HemmatiS18,DBLP:conf/icmla/LachmannSNSS16}, but they also rely on manually labeled data.
None of them provides tools to guide testers to test applications more effectively, which is studied in this work.
Crowdtesting is a newly developed manual testing schema~\cite{wang2020context,wang2019iSENSE,wang2021context} in which software development companies release test tasks through a crowd platform~\cite{Amazon,Applause}. Crowd workers conduct testing according to the description of testing tasks. However, researches showed that crowd workers always test the same function or repeated interfaces~\cite{xie2017cocoon,wang2019images}, which leads to the waste of time and efforts of developers and testers. 
To avoid those repetitive operations, Zipt et al.~\cite{deka2017zipt} and Yan et al.~\cite{chen2020improving} track each testers' behavior and aggregate them for reminding testers.
In the context of GUI testing, a common method is to remove duplicate items from the list of test cases through post-event result analysis~\cite{wang2019images}.
Other researchers have proposed incentive-based methods to reward testers who have found previously undetected cases~\cite{attenberg2011beat}.
Although these works can avoid duplicate efforts, they cannot help cover more activities or unrevealed bugs within the application. 
The testers may still miss some important application functionalities during testing.
In addition to reminding users with testing repetition, our approach can guide the human testers during the manual exploration for achieving higher activity coverage and potentially uncovering more bugs.

\textbf{Manual Test Assistant Technology.}
To give full play to the advantages of manual testing, previous studies utilized test assistant related technologies to improve the quality of manual testing~\cite{frisson2016inspectorwidget,wang2017clickstream,chen2021cocapture}.
SwiftHa\cite{choi2013guided} learned the models of Android applications and used them to discover unexplored states. Monkeylab~\cite{linares2015mining} modeled user event interaction sequences on Android applications to generate new test cases. Polaris~\cite{mao2017crowd} simulated user interaction patterns learned from user behavior on Android applications, and then applied  this simulation to different applications. Rico~\cite{deka2017rico} proposed a hybrid method to record the application tracking of group workers for the first time, and then continued to explode programmatically to achieve a wider state space in the application. Patina~\cite{matejka2013patina} proposed an application-independent system for collecting and visualizing software application usage data. 
These methods combine the advantages of human intelligence and machine intelligence, so that the test cases are realistic and the testing tasks are scalable~\cite{ponsard2016anchored,fourney2015web}. However, the challenges of test duplication and incompleteness still remain. Chen et al.~\cite{chen2020improving} proposed GUI level guidance on a web app to solve the problem of repeated testing by testers.
There are three aspects distinguishing their work from ours. First, our approach targets at the Android apps, which has gained increasing popularity in people's daily life. Second, besides telling the testers where have been explored to avoid repetition, our approach can also guide the testers in exploring more UI pages to find more bugs. Third, we design an effective path exploration  algorithm that helps to plan the optimized testing steps to void repetitions more effectively.

\textbf{State Transition Graph Extraction.}
In the Android app, an state (e.g., activity) usually corresponds to an interface. 
The activity transition graph (ATG) reflects the relationship between activities. 
As the foundation of the downstream software testing task, ATG is usually used to analyze the  relationship between activities in Android app.
Zhang et al.~\cite{zhang2018launch} adopt the launch-mode-aware context-sensitive activity transition analysis method to extract ATG. 
Yang et al.~\cite{yang2018static} proposed a method of extracting static window transition graph (WTG) for Android based on window stack.
Yet these methods can only obtain the coarse-grained ATG.
Different from them, this work aims to obtain a fine-grained state transition graph (STG) by combining both the static program analysis and dynamic exploration.
Besides the nodes and edges in the graph, we enrich the STG by annotating the trigger actions along with the edge.

\section{motivational study}
\label{sec_motivation}
To understand the bugs during manual exploratory testing~\cite{itkonen2007defect, itkonen2005exploratory}, we carry out an empirical study on observing testers' behavior.
We randomly select 85 apps from F-Droid (the largest repository of open-source Android apps) according to the number of downloads, including 17 categories (e.g., connectivity, games, internet, money, reading, education, health) with 5 latest released apps in each category. 
Note that, to ensure the popularity of the experimental apps, these selected apps are also published in Google Play.
Activity number in each app ranges from 10 to 30, and more detailed information of apps can be seen in our website\textsuperscript{\ref{github}}.
We recruit 10 testers, all of whom major in computer science with more than 3 years of app testing experience.
Six of them are from industry with practical working experience, while the other four are master students.
Since the target users of this paper are testers who are not familiar with the application they are testing, none of the participants we selected have used the above apps.

Each tester independently completes the exploratory test of all apps, and the maximum test time of each application is set up 10 minutes. 
During the experiment, we ask testers to record their screen to ensure the validity of their testing and for further analysis.
After the experiment, we carry out an informal interview with these 10 testers to further understand what they think about the testing.
The following observations are found: 

\textbf{1) Low activity coverage vs. High confidence.} 
The activity coverage rate of exploratory testing is only 57\% on average, indicating that it is hard for human testers to cover all activities or functionalities of an application when conducting exploratory testing. 
As seen in Fig~\ref{fig:example_motivation} (The MoneyTracker app), some functionalities are rarely explored such as ``About'' in the Setting page and ``change the date'' in the Main page by one user.
Despite the low coverage, most of the participants are confident that they cover most functionalities of apps. 
Such blind confidence may significantly hurt the quality of testing. 

\textbf{2) Repeated visiting vs. Unawareness.}
85\% of the testers repeatedly visit the same page more than 10 times (e.g., ``Account'' and ``Report'' page in Fig~\ref{fig:example_motivation}), and 65\% of the testers are trapped into the loop for more than 3 minutes according to our observation of the video recording.
We also find that some testers hesitated for a long time on one page, without knowing the next step.
Some of them mention that they visit some pages repeatedly, as they could not remember which page has been visited or which action can trigger a new page.
However, there are still a large portion of participants that are  unaware of their repeated testing within the app.

In summary, the low activity coverage of manual testing confirms the necessity of guidance during the testing process.
The high confidence of testing and the unawareness of the repetition further indicates this practical need.
In addition, the repeated visiting phenomenon motivates us in developing a state transition graph to record what has been explored and guide human testers to efficiently explore the uncover states in order to facilitate the exploratory app testing.

\begin{figure*}[htb]
\centering
\vspace{0.2in}
\includegraphics[width=17.7cm]{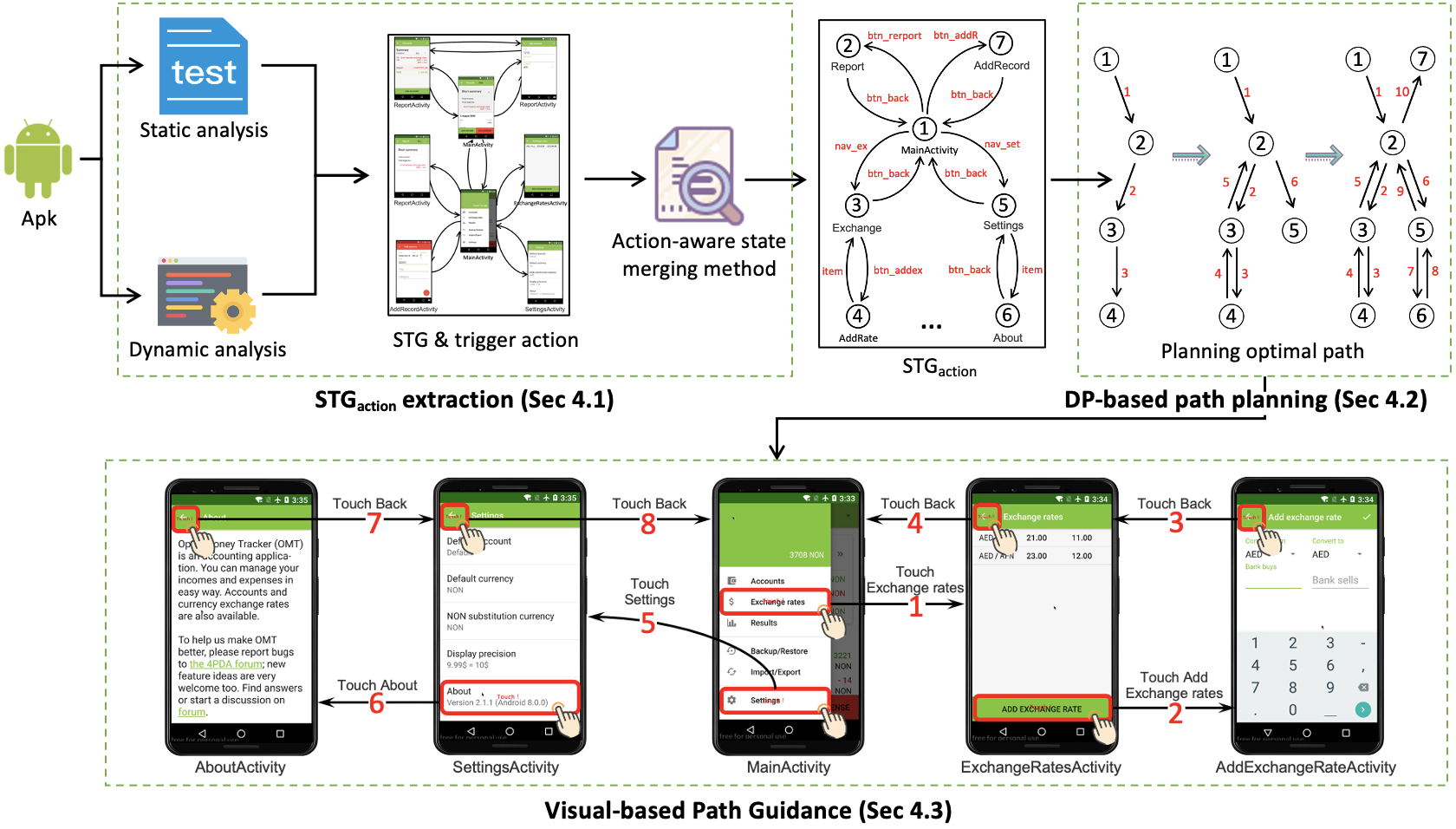}
\vspace{-0.1in}
\caption{Overview of {\tool}}
\label{fig:overview}
\vspace{-0.1in}
\end{figure*}

\section{Approach}
\label{sec_approach}
This paper proposes {\tool} to navigate human testers in exploring apps to avoid missing functionalities or making repeated exploration steps.
Fig. \ref{fig:overview} presents the overview of {\tool}, which consists of three main components. 
\textbf{\textit{First}}, given the app Android package, the $STG_{action}$ extraction component combines both static analysis and dynamic exploration to extract the state transition graph (STG) and its trigger actions between states (Section \ref{subsec_approach_ATG_extraction}). 
We also design a context-aware state merging method to merge near-duplicate states by considering the current state and the adjacent states. 
\textbf{\textit{Second}}, based on the extracted $STG_{action}$, the DP-based path planning component plans the exploration path which aims at covering all the states of the app with few repeated exploration steps (Section \ref{subsec_approach_Path}).
\textbf{\textit{Third}}, with the planned path, the visual-based path guidance component utilizes the visual augmentation technology to guide users' testing (Section \ref{subsec_approach_Visual}).

\subsection{$STG_{action}$ Extraction}
\label{subsec_approach_ATG_extraction}

We first extract the state transition graph (STG), and then enrich it with trigger actions between the states to construct $STG_{action}$, which serves as the basis in guiding the users exploring the app.
Both static analysis and dynamic explorations are used to ensure that accurate states and trigger actions are captured. 

In our method, $STG_{action}$ is defined as a graph $G <N, E>$ with node $N \in state$ and edge $E \in action$.

\textbf{State}: Previous automated testing works adopt different standards (e.g., activity, UI page) to abstract the $state$ of an application~\cite{pan2020reinforcement,adamo2018reinforcement,azim2013targeted,baek2016automated}. 
Research shows that more fine-grained $state$ abstraction may lead to higher testing coverage~\cite{baek2016automated}.
Inspired by app GUI testing~\cite{pan2020reinforcement, machiry2013dynodroid, Monkey}, we regard each unique UI page as one $state$ and represent it by i.e., represented by UI components hierarchy tree, in which non-leaf nodes as layout components (e.g., LinearLayout, Framelayout) and leaf nodes as executable components (e.g., button);
Each $activity$ may have multiple $state$, that is, $N \in state \in activity$.

\textbf{Action}: Action is the trigger that results in $state$ transition, which can be expressed as $E = ID, \ E \in action$. Where, $ID$ is the identity of GUI component which receives the action.

\subsubsection{\textbf{Static STG Extraction and Trigger Action Detection}}

In an Android application, activities can be started by invoking.
For example, the $StartActivity(intent)$ is an inter-component communication (ICC) call~\cite{yanmultiple}, passing an intent that describes the activity to be launched~\cite{zhang2018launch}.
By analyzing the ICC, the activity transition graph (ATG)~\cite{azim2013targeted,mirzaei2016reducing} can be extracted.
That ATG can be regarded as an initial $STG_{action}$.

In detail, the target activity of ICC call is determined by querying the pointed-to values in the fields of an intent object. For example, $StartActivity(intent)$ determines the target activity to be started. 
By matching the parameter in $intent()$ method with the parameter in \textit{AndroidManifest.xml} file, we obtain the transition between activities and build the initial ATG.

To guide the tester in manual testing, it is necessary to further extract detailed action (e.g., clicking a specific button) that triggers the activity transition. 
Given an application, the trigger action can be extracted by analyzing the call of $intent()$. 
We first transform the application source code to an abstract syntax tree (AST), and then traverse AST to locate $intent()$ method.
There are two detailed types of it.
First, if the $setOnClickListener()$ of a component (such as a button) directly calls the $intent()$ method, we will directly obtain the component name that calls the intent method.
We traverse the AST again to obtain the corresponding trigger component which is represented by a certain ID (e.g., $R.id.btn\_back$).
Second, if the component calls the $intent()$ method when calling other methods, we locate the component which calls $intent()$ by traversing the method name $launchHome()$.
After obtaining the component ID corresponding to the trigger action, the edge connecting two activities is determined, and we build the $ATG_{action}$ (i.e., coarse-grained $STG_{action}$) accordingly.

\subsubsection{\textbf{Dynamic STG Extraction and Trigger Action Detection}}
\label{subsubsec_approach_Trigger_action_detection_dynamic}

Some states and trigger actions, especially those in dynamic or mixed layout (such as dynamic rendering menu, list, navigation bar, appwidget), are difficult to be obtained by static analysis~\cite{wang2016unsoundness,yang2018static}.
Instead, it is easy to be captured with dynamic GUI rendering.

Therefore, to enrich the $STG_{action}$ extracted in the above section, we further adopt dynamic exploration to detect more fine-grained states and actions.
In detail, by leveraging the idea of dynamic app GUI testing~\cite{li2017droidbot,cai2020fastbot,Monkey,su2017guided}, we adopt an app explorer~\cite{li2017droidbot} to automatically explore the pages within an application through interacting with apps using random actions e.g., clicking, scrolling, and filling in text. 

During the exploration, we record both $state$ and trigger action $action$ between states. 
Each $state$ corresponds to the detailed view hierarchy file (XML file) of one Android application page.
Each $action$ corresponds to the component ID that triggers the state transition.
For the case that the ID of some dynamic rendering components cannot be retrieved in the run-time GUI hierarchy, we analyze the `text' corresponding to the component in the view hierarchy file and treat the text content as its ID in $STG_{action}$. 
If the component has no `text', we treat the coordinates of the component as the component ID.
We then combine the $STG_{action}$ extracted from static analysis and dynamic exploration into one graph.

\subsubsection{\textbf{Context-aware State Merging}}
\label{sec:stateMerging}
\begin{figure}[htb]
\centering
\vspace{-0.05in}
\includegraphics[width=8.3cm]{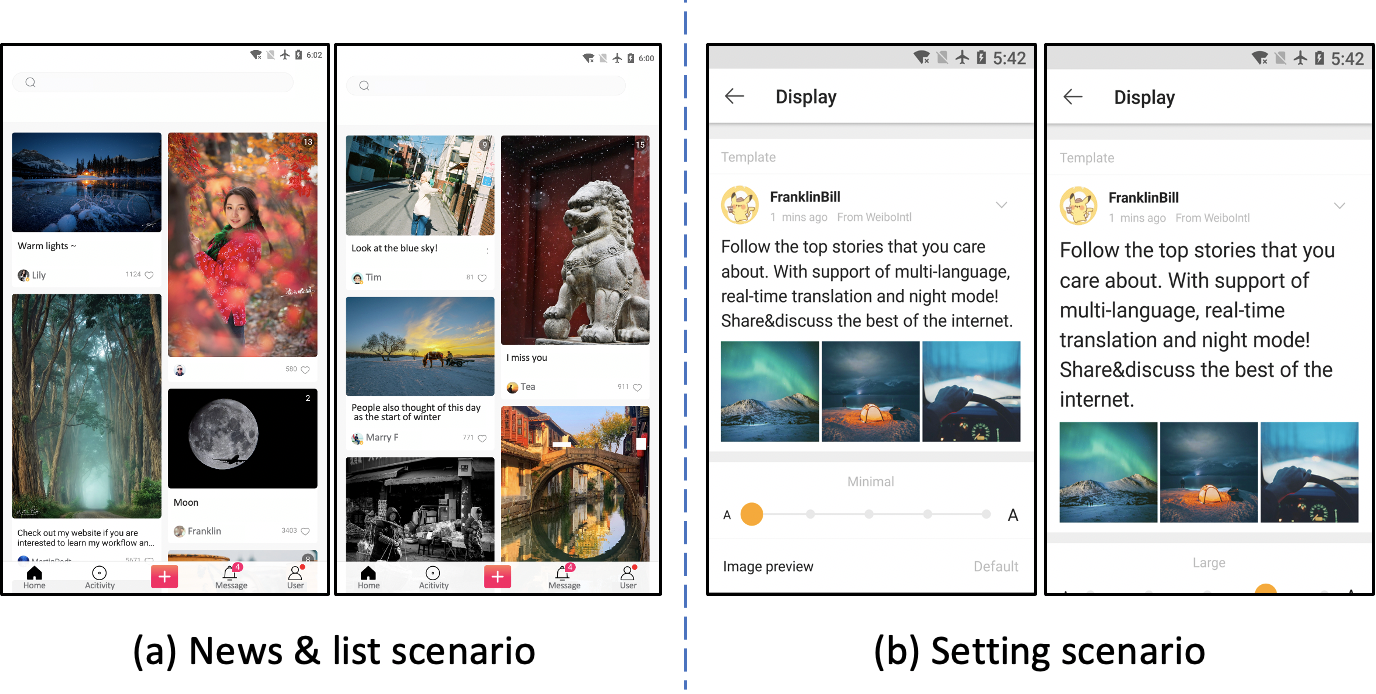}
\vspace{-0.05in}
\caption{Example of failed cases with existing approaches.}
\label{fig:scenario-example}
\vspace{-0.05in}
\end{figure}

Through static and dynamic analysis, we get $STG_{action}$ which is composed of a large number of $states$ and $actions$, in which some of them are duplicates~\cite{yandrapally2020near,fetterly2003evolution,manku2007detecting}.
To avoid state explosion, we develop a context-aware based approach to merge duplicate states.
Existing works either consider run-time GUI hierarchy~\cite{su2017guided, pan2020reinforcement} or visual features~\cite{yandrapally2020near} to remove near-duplicate states.
Nevertheless, some near-duplicate states can still be missed by these approaches, or some non-duplicate states might be wrongly detected.
For example, streaming recommendation with different content in different time/users in Fig~\ref{fig:scenario-example} (a) can hardly be correctly detected by visual features, while the same content in different font size settings in Fig~\ref{fig:scenario-example} (b) could not be detected by run-time GUI hierarchy.

Therefore, to overcome those drawbacks, we consider not only the information within a certain state but also its context i.e., adjacent states, for state merging.
Given $STG_{action}$, we first merge the states with the same GUI run-time hierarchy (XML file from $ADBdump$ command) without considering detailed content (e.g., text or image) which may change dramatically.
After that, we further merge states with similar GUI hierarchy by checking whether their $n-1$ $state$ (i.e., the previous state which transits to the current state) and $n+1$ $state$ (i.e., the next state to which current state transits) are similar.
By referring to existing studies~\cite{pan2020reinforcement,yandrapally2020near,li2017droidbot} and combining with our empirical observations, we set the threshold of similarity as 80\%.
Following these two operations, we finish the state merging.

Although we construct a $STG_{action}$ for supporting the real-time path recommendation during testers' manual exploration, there is still potentially a gap between states stored in our $STG_{action}$ and states in the real-world testing environment.
As testers' input (e.g., text) or exploration path may be different from that in our static analysis and dynamic exploration, we need to map the live state to those in our $STG_{action}$ which may be slightly different.
We adopt the same approach described above for state mapping.

\subsection{DP-based Path Planning}
\label{subsec_approach_Path}

With the $STG_{action}$ obtained in the previous section, we need to plan a path that can cover all the nodes (i.e., states) with few repeated steps, so as to serve as the basis for the follow-up testing guidance.
To achieve this, we use a dynamic programming algorithm to derive the shortest path between each pair of nodes, and plan the path.

\subsubsection{\textbf{Formalization of Planning Path}}
\label{subsubsec_approach_Formal}

We formulate the path planning as a dynamic programming problem, and represent it by a 4-tuple: $\bm{< G, d, V, DP>}$.

$\bm{G}$: \textbf{Graph}. The $STG_{action}$ ($\bm{G <N, E>}$) obtained in the previous section, where $\bm{N}$ is the set of nodes \textit{(i.e., states)}, and $\bm{E}$ is the set of edges \textit{(i.e., triggered events)}. 

$\bm{d}$: \textbf{Distance}. $\bm{d_{ij}}$ is the shortest distance between the state $i$ and the state $j$.

$\bm{V}$: \textbf{Visit status}. $\bm{V}$ is the visit status of the current node, represented by binary numbers. 0 is not visited and 1 is visited.

$\bm{DP}$: \textbf{Dynamic programming}. $\bm{DP_{jV}}$ is the shortest distance from the current state $i$ to state $j$ in visit status $\bm{V}$. Since $\bm{V}$ is a binary number, $\bm{DP_{i(V\land(1 \ll (j-1))}}$ is the distance of reaching state $i$ without accessing other states ($\ll$ is the bitwise operator).

Under the above formalization, to solve the path planning problem is to optimize the following two equations:

$$\left\{
\begin{array}{l} 
d_{ij} = min(d_{ij}, d_{ik} + d_{kj})\\
DP_{jV} = min(DP_{jV}, DP_{i(V\land(1 \ll (j-1))} + d_{ij})
\end{array}
\right.
$$

\subsubsection{\textbf{Planning Strategies}}
\label{subsubsec_approach_Exploration}
The dynamic programming algorithm is shown in Algorithm 1.
Its general idea is to use multi-stage optimal decision-making, where each decision depends on the current visit status, and then cause the visit status to transfer.
In detail, the algorithm first traverses the graph $STG_{action}$ to obtain the set of nodes $N$ and edges $E$ (line 1-2). 
It then employs Floyd algorithm~\cite{mirino2017best} to calculate the shortest path between each pair of nodes $d[i][j]$ , and record the node sequence for each shortest path in $NodeSequence$ (line 3-10).

\begin{algorithm} 
    \label{alg:PG} 
    \caption{DP-based Path planning algorithm}  
    \KwIn{
        $STG_{action}$: $STG_{action}$ graph\;
        }
    \KwOut{
        $path$: Recommended path to users\;
        }
    $N \gets getNode(STG_{action})$\;
    $E \gets getEdge(STG_{action})$\;
    
    \For{each $k\in n$}
    {
        \For{each $i\in n$}
        {
            \For{each $j\in n$}
            {
                \If{$d[i][j] > (d[i][k] + d[k][j])$}
                {
                    $NodeSequence[i,j] = k$\;
                    \textcolor{gray}{//Record nodes sequence of shortest path.}
                }
                $d[i][j] = min(d[i][j], d[i][k] + d[k][j])$\;
                \textcolor{gray}{//Calculate the shortest distance between nodes.}
            }
        }
    }
    
    $DP[n][n] \gets inf$\;
    $bit \gets$ $(1\ll n)$\;
    $DP[0][1] \gets$ $0$\;
    \For{each $V\in bit$}
    {
        \For{each $i\in n$}
        {
            \If{$V \ \& \ (1\ll i)$}
            {
                \For{each $j\in n$}
                {
                    \If{not $(V \& (1\ll j)$ and $d[i][j]$ != $inf$} 
                    {
                        \If{$DP[j][V] > DP[i][V\land(1 \ll (j-1))] + d[i][j]$}
                        {
                            $VisitStatus[i,(V \land (1 \ll j))] = (j,V)$\;
                            \textcolor{gray}{//Update node visit status.}
                        }
                        $DP[j][V]=min(DP[j][V],DP[i][V\land(1 \ll (j-1))] + d[i][j])$\;
                    }
                }
            }
        }
    }
    $minn \gets inf$\;
    \For{each $node \in n$}
    {
        \If{$minn > DP[node][bit-1]$}
        {
            $path\gets$ $getpath(NodeSequence,VisitStatus,node,bit-1)$\;
            \textcolor{gray}{//Get the planned path according to the node status.}
        }
        $minn = min(minn, dp[node][bit - 1])$\;
    }
    
    return $path$\;  
\end{algorithm}

Based on the shortest path information, the algorithm then begins the path planning. 
Specifically, it maintains a buffer to store the visit status $V$, and gives priority to the nodes that have not been visited. 
Suppose the exploration is currently at node $i$, the algorithm will judge whether the visit status $DP[j][V]$ of node $j$ is visited; if not, it finds a shortest path $d[i][j]$ between node $i$ and node $j$, and update the node visit status $VisitStatus$ (line 11-22).
Finally, by calculating the shortest distance $minn$ from the start node to the end node, the algorithm can get the end node $node$ and its visit status $bit-1$. 
According to the nodes sequence information of the shortest path in $NodeSequence$ and the visit status information in $VisitStatus$, each intermediate node is backtracked from the tail node $node$ to derive the visit order of nodes. 
After all nodes in the graph are visited, the algorithm can recommend the planned path for testers to explore (line 23-29).

Note that if the tester does not follow the path recommended by our approach in the process of exploration, we would record the state when he/she changes the path.
Then according to the path that has been explored, {\tool} will recalculate the path by running Algorithm 1 and take current state as the starting point.

\subsection{Visual-based Path Guidance}
\label{subsec_approach_Visual}

We further implement the planned path into {\tool} for guiding testers in testing mobile apps.
It can suggest the next operation step by step in the user interface to help the testers covering the unexplored pages and reducing replication explorations. 
Specifically, we augment the run-time GUI with visual hint moves.

{\tool} uses the Android debug bridge (adb)~\cite{Adb} command to start the app that testers need to test. 
To run an Android app, the source code is compiled and the mobile devices use rendering to realize the display of Android UI on the screen.
During the tester's exploration, {\tool} obtains the run-time information of the current interface including the state information and existing components within the current page on the backend. 
Specifically, {\tool} can obtain the view hierarchy file corresponding to the current UI page (state) through ``uiautomator dump'' of the Android Debug Bridge (adb) command~\cite{Android}. The view hierarchy file includes the component information (coordinate information, ID, component type, text description, etc.), and the layout information on the current state after rendering~\cite{Viewhierachy}.
Given the state information of the current page, we search the obtained $STG_{action}$ on the fly, find the next state on the planned path, and highlight the corresponding actions which can trigger that state in the page.

\begin{figure}[htb]
\centering
\vspace{-0.05in}
\includegraphics[width=8.0cm]{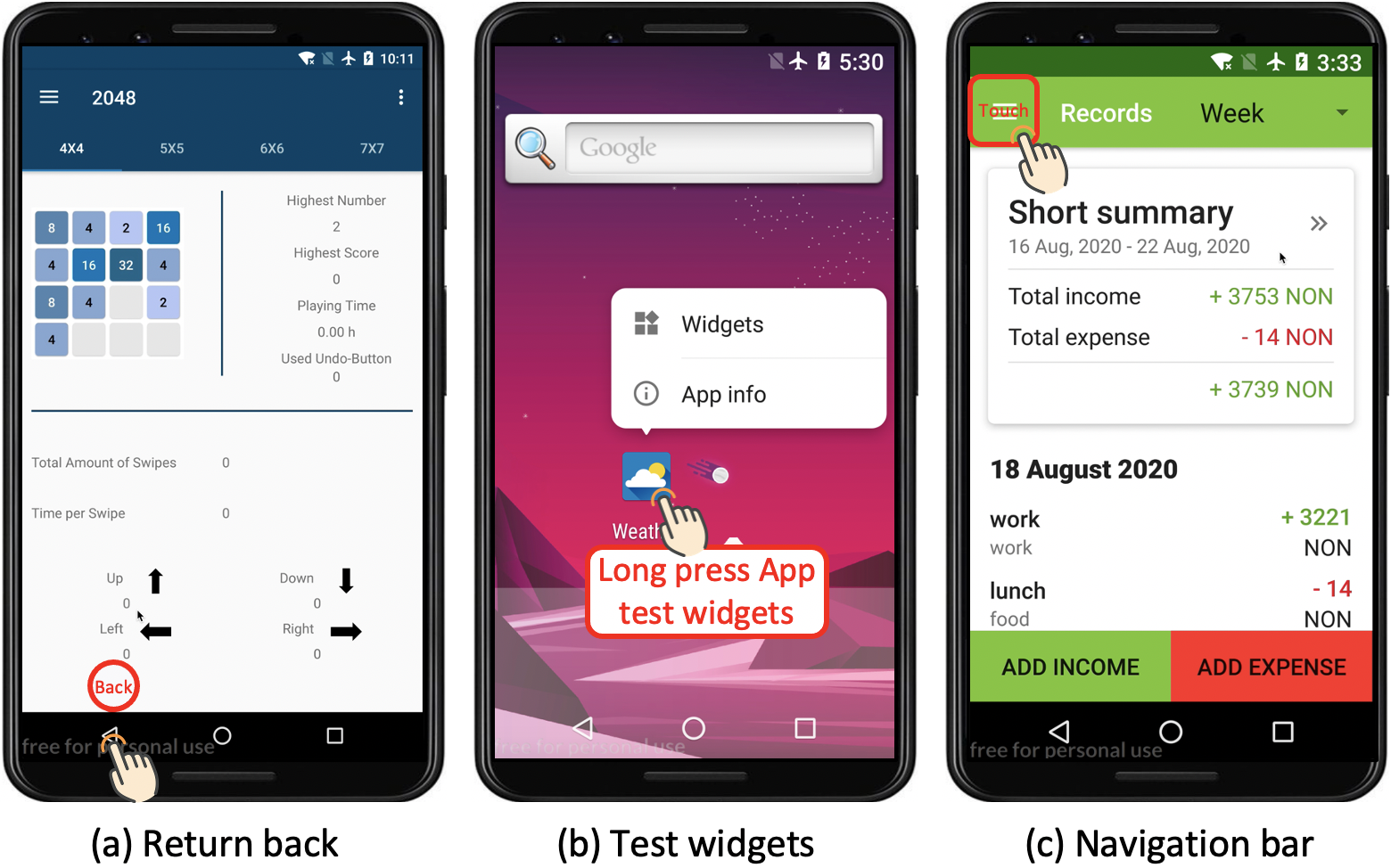}
\vspace{-0.1in}
\caption{Highlighted suggested moves from {\tool}.}
\label{fig:3-state}
\vspace{-0.1in}
\end{figure}

\begin{figure*}[htb]
\centering
\hspace{-0.2in}
\includegraphics[width=18.0cm]{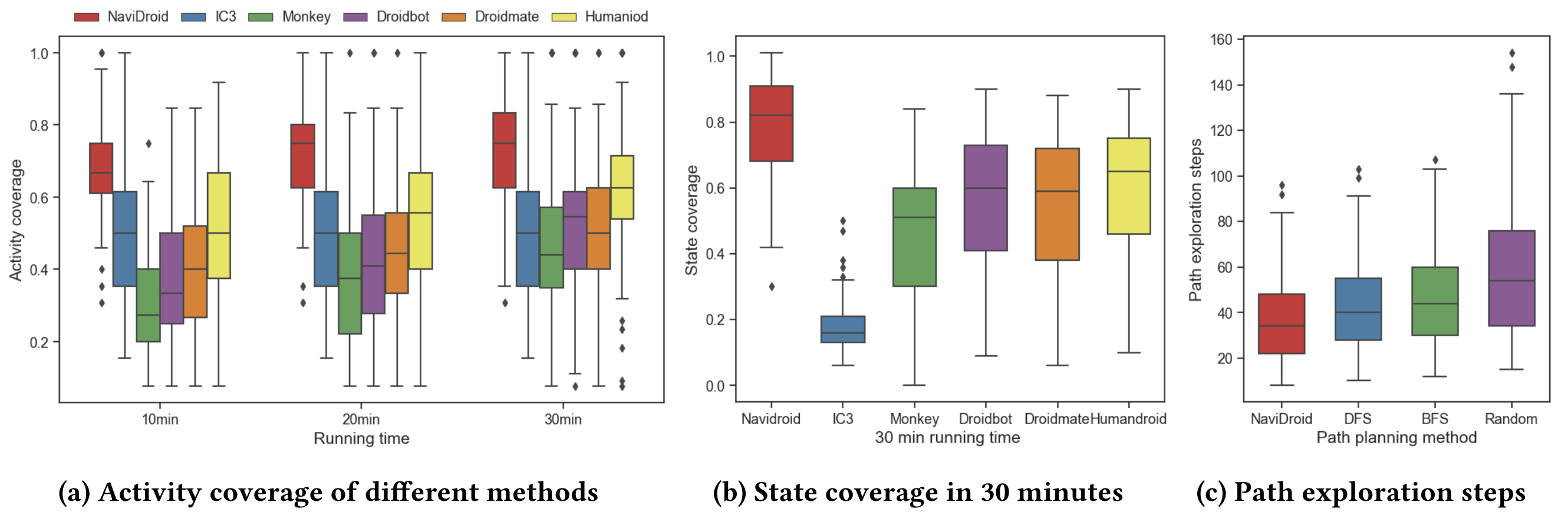}
\vspace{-0.1in}
\caption{Result of effectiveness evaluation}
\label{fig:RQ1-all}
\vspace{-0.15in}
\end{figure*}

We adopt the Android floating window ~\cite{chen2020improving, li2016visualizing} for visualizing the hint moves.
It is a mobile window, which floats on the top of an app.
As the Android interface drawing is realized through the services of $WindowManager$, which can add the floating window control to the screen through the $AddView()$ method.
After getting the trigger action from $STG_{action}$, our {\tool} uses the adb command (``\textit{adb shell uiautomator dump --compressed}'') to get the coordinates and sizes of the component from the view hierarchy file~\cite{Viewhierachy} that testers need to operate. 
The system runs the floating window service in the backend, and sets the size and coordinates of the floating window (the rectangular box as shown in Figure \ref{fig:3-state}) to make it in the same position and suitable size as the component and floating on the component, so as to guide the testers to explore the application interface.

There are three main types of trigger actions, i.e., returning back, long-pressing test widgets, and clicking a component(e.g., button, navigation bar) . 
The first action is easily visualized by a bounding box floating on the button as the hint.
We present the last three cases in Fig \ref{fig:3-state} to facilitate understanding.
If there is no back button in the current interface and the id is `touch\_back' as in Fig \ref{fig:3-state} (a), {\tool} will suggest the tester with ``back'' action above the back key, otherwise, the tool will directly highlight the back button.
As shown in Fig \ref{fig:3-state} (b), 
for the ``appwidget activity'', the tester will be suggested to operate ``long press app test widgets'' through the floating window when exploring the app. 
As shown in Fig \ref{fig:3-state} (c), for a button, navigation bar or fragment, the tester is suggested to click the UI component.

\section{Effectiveness Evaluation}
\label{sec_Effectiveness}



{\tool} consists of STG extraction algorithm, dynamic programming algorithm and visual guidance. Therefore, we first evaluate the algorithm performance of the {\tool} through an automated method.
We evaluate the effectiveness of {\tool} from the points view of $STG_{action}$ extraction component (Section \ref{subsec_approach_ATG_extraction}) and DP-based path planning component (Section \ref{subsec_approach_Path}) respectively. 
For $STG_{action}$ extraction, we compare the activity coverage and state coverage  with 5 baseline methods to demonstrate its advantage (details are in Section \ref{subsec_experiment_baseline}).
For path planning, we run {\tool} and record the number of exploration steps, and then compare the path exploration efficiency of the {\tool} with three baselines.

\subsection{Experiment Setup}
\label{subsec_experiment_Setup}

\subsubsection{\textbf{Dataset and Experiment Procedures.} }
\label{subsec_experiment_dataset}

We use the 85 open-source mobile applications, as demonstrated in Section \ref{sec_motivation}, for experiment, more detailed information of apps can be seen in our website\textsuperscript{\ref{github}}.

For activity coverage, we collect all the activities defined in each app from AndroidManifest.xml following existing studies~\cite{chen2019storydroid,li2017droidbot,cai2020fastbot}, and compare the percentage of the explored activities by running {\tool} for 30 minutes.
Note that there may be multiple states in one activity, so we also use state coverage~\cite{su2017guided} for evaluation. Since the app state cannot be obtained directly from Android files~\cite{su2017guided,yang2018static}, we invite two testers with more than five years of testing experience to manually label the states of 85 apps for the effectiveness evaluation.
They conduct the manual labelling separately, and discuss the difference until a common consensus is reached.
For path planning, we measure the number of exploration steps which is widely used~\cite{perez2018learning,shwail2013probabilistic}.

\subsubsection{\textbf{Baselines}}
\label{subsec_experiment_baseline}

For activity and state coverage, to further demonstrate the advantage of {\tool}, we compare it with 5 common-used and state-of-the-art baselines.
There are one static analysis based tool and four automated testing tools, i.e.,  IC3~\cite{octeau2015composite}, Monkey~\cite{Monkey}, Droidbot~\cite{li2017droidbot}, DroidMate~\cite{borges2018droidmate} and Humanoid~\cite{li2019humanoid}. The above automated test method is selected to evaluate the effect of {\tool}'s algorithm on activity coverage.
We run {\tool} and these baselines on an Android virtual machine of Google Nexus 6 with Android 6.0 OS. 
We use the default configuration settings for each tool, record the activity and state coverage by running the test for 30 minutes.
The experiment is repeated three times for robust consideration, and the average performance is used.
For path planning, based on the extracted $STG_{action}$, we choose 3 baselines of path exploration, i.e.,  Depth-First-Search~\cite{shwail2013probabilistic,tarjan1972depth}, Breadth-First-Search~\cite{beamer2012direction}, Random Exploration~\cite{Monkey,borges2018droidmate}.
And then we record the number of exploration steps.

\subsection{Results and Analysis}
\label{subsec_results}

\subsubsection{\textbf{Activity and State Coverage}}
\label{sec_results_RQ1}

Fig \ref{fig:RQ1-all}(a) shows the activity coverage of {\tool} and the baselines.
{\tool} achieves a median activity coverage of 0.74 with the range from 0.5 to 1.0 across 85 mobile apps.
Fig \ref{fig:RQ1-all}(b) shows the state coverage of {\tool} and the baselines in 30 minutes (Because the overall trend of state coverage and activity coverage is similar, we only give the state coverage in 30 minutes due to space constraints, and we upload the complete data to our website.).
{\tool} achieves a median state coverage of 0.81 with the range from 0.6 to 1.0 across 85 mobile apps.
This indicates the effectiveness of {\tool} in covering most of the activities and states so as to provide a full viewpoint for guiding testers in exploring the app.

{\tool} is 12\% (0.74 vs. 0.62) higher even compared with the best baseline (Humanoid) in activity coverage (with 30 minutes running time).
This is because these baseline approaches rely on random user actions to explore the app, and the activity coverage could not be ensured.  

Fig \ref{fig:RQ1-all}(a) also shows the coverage variation in terms of different time intervals.
We can see that {\tool} can quickly achieve a high coverage within 10 minutes. 
In contrast, automated testing tools need far more time to achieve their own optimal coverage. 
This is mainly because our {\tool} can get a certain number of activities through static analysis, and combining them with the states obtained from dynamic analysis can achieve a superior activity coverage.

Although static analysis based baseline IC3 is not sensitive to the testing time, it misses two types of activities compared with our {\tool}.
In detail, it is unable to analyze the dynamic layout and mixed layout, and cannot handle the conversion of Android system event callback. 
For the automated testing tools, the activity coverage increases along with the testing time.
However, even given 30 minutes for testing one application, they can not reach as high activity coverage as that of our approach.
That is because some apps require login or complicated input for triggering a certain activity which cannot be achieved by random test case generation.
It also indicates the necessity of manual testing and the importance of the development of our {\tool} for assisting the manual testing.

We further examine the uncovered activities and states by our {\tool} and summarize the following two reasons. 
First, some apps contain AppWidget activity, which is difficult to obtain the corresponding startup button.
As running AppWidget activity requires a series of operations (i.e., exit the application to the home page of the device, hold down the application icon on the home page, then move the AppWidget to the device home page), obtaining the coordinates of the AppWidget is quite difficult.
Second, there are some navigation bars or menus with numerical item ID (i.e., 1,2,3,...,N), rather than the meaningful descriptive one, and using the switch case method to call $intent()$.
These numerical item IDs cannot be accurately obtained, and the trigger action cannot be extracted.

\subsubsection{\textbf{Path Exploration}}
\label{sec_results_RQ1_2}

Fig \ref{fig:RQ1-all}(c) shows the path exploration steps of our {\tool} and the three baselines.
In 85 mobile applications, the average number of path exploration steps of {\tool} is 36, outperforming the baselines.
{\tool} saves 20\% (45 vs. 36), 23\% (47 vs. 36) and 42\% (60 vs. 35) steps compared with DFS, BFS and random exploration respectively (Each app is randomly explored once.).
Since there are many leaf nodes and ring structures in $STG_{action}$, the three baseline approaches can occasionally fall into the ring structures, resulting in repeated operations. 
These baselines employ heuristic-driven strategies, and are far from achieving the globally optimal solution.
By comparison, our proposed approach considers the whole graph during path planning, and can achieve more optimal exploration outcomes with fewer repeated steps.

\section{Usefulness Evaluation}
\label{sec_Usefulness}
To evaluate our {\tool}, we also conduct a user study to demonstrate its usefulness in the real-world practice of manual testing. 
Our goal is to examine: (1) whether {\tool} can effectively help explore the functionality of the application? (2) whether {\tool} can help users find more bugs? (3) whether {\tool} can save users' testing time?

\subsection{\textbf{Dataset of User Study}}
\label{sec_GUI_Testing_Tasks}
We begin with the 85 apps from F-Droid (also in Google Play) described in Section \ref{sec_motivation}.
To get realistic tasks for testing, we recruit an independent professional tester with four years of testing experience of Android app from Tencent. 
We ask him randomly choose apps, and find UI display bugs from each app.
Note that, to ensure that testers can run the app smoothly, there are no functional bugs such as crashes in apps, and the UI display bugs will not cause any app crash. 
We end up with 20 apps with 30 UI display bugs (each app with at least one bug), and use them for the final evaluation, with details in Table \ref{tab:User study}.

\subsection{Participants Recruitment}
\label{sec_Participants}
We recruit 32 testers to participate in the experiment (different from Section \ref{sec_motivation}) from a crowdtesting platform TestIn.
According to the background survey, all participants graduate with a computer science degree. They all have more than two years of app testing experience and practical work experience in the industry. Every participant receives a \$50 shopping card as a reward after the experiment. At the beginning of the test, participants are asked to watch a short tutorial video and familiarize themselves with the apps. We also conduct a follow-up survey among the participants regarding their experiment experience.

The study involves two groups of 32 participants: the experimental group from P1 to P16 who test the mobile apps guided by our {\tool}, and the control group from P17 to P32 who conduct the testing without any assistant.
Each pair of participants $\langle$ Px, P(x+16) $\rangle$ have comparable app testing experience to ensure that the experimental group has similar expertise and capability to the control group in total. 

\begin{table*}[htb]
\renewcommand\arraystretch{1.4} 
\vspace{0.1in}
\caption{The comparison of the experiment and control group.}
\label{tab:User study}
\centering
\footnotesize
\begin{tabular}{p{0.2cm}<{\centering}p{1.2cm}<{\centering}p{0.8cm}<{\centering}|p{0.8cm}<{\centering}p{1.3cm}<{\centering}|p{0.8cm}<{\centering}p{1.3cm}<{\centering}|p{0.8cm}<{\centering}p{1.3cm}<{\centering}|p{0.8cm}<{\centering}p{1.3cm}<{\centering}}
\hline
\multicolumn{3}{c|}{\textbf{Basic information}} & \multicolumn{2}{c|}{\textbf{State coverage}} & \multicolumn{2}{c|}{\textbf{Activity coverage}} & \multicolumn{2}{c|}{\textbf{Time (min)}} & \multicolumn{2}{c}{\textbf{\# Bugs}}  \cr\cline{1-11}
 \textbf{id} & \textbf{App} & \textbf{Categ} & \textbf{control} & \textbf{experiment} & \textbf{control} & \textbf{experiment} & \textbf{control} & \textbf{experiment} & \textbf{control} & \textbf{experiment}\\
\hline
1 & PHealth & Health & 0.51  & 0.73  & 0.53  & 0.78  & 4.50  & 2.08  & 0.69  & 1.94 \\ 
2 & Weather & Internet & 0.46  & 0.72  & 0.52  & 0.80  & 4.51  & 2.67  & 0.81  & 2.56 \\ 
3 & Contact & Phone & 0.50  & 0.73  & 0.57  & 0.73  & 5.26  & 2.91  & 0.81  & 1.81 \\ 
4 & MoneyTK & Finance & 0.39  & 0.68  & 0.45  & 0.82  & 7.79  & 4.31  & 0.94  & 2.00 \\ 
5 & FoodFacts & Health & 0.53  & 0.78  & 0.55  & 0.83  & 8.28  & 6.71  & 1.44  & 2.63 \\ 
6 & GPSTest & Navig & 0.44  & 0.71  & 0.53  & 0.90  & 5.92  & 4.18  & 0.50  & 1.00 \\ 
7 & PSStore & Security & 0.44  & 0.68  & 0.49  & 0.76  & 8.18  & 6.68  & 0.38  & 0.94 \\ 
8 & NewPipe & Media & 0.43  & 0.74  & 0.52  & 0.81  & 6.46  & 5.09  & 0.75  & 1.63 \\ 
9 & WallETH & Finance & 0.48  & 0.80  & 0.49  & 0.84  & 8.52  & 7.76  & 0.38  & 0.94 \\ 
10 & Transistor & Media & 0.38  & 0.76  & 0.40  & 0.82  & 5.33  & 3.65  & 0.31  & 0.94 \\ 
11 & Democracy & News & 0.42  & 0.75  & 0.45  & 0.81  & 6.68  & 5.31  & 0.25  & 0.94 \\ 
12 & Metrodroid & Navig & 0.31  & 0.60  & 0.39  & 0.72  & 7.37  & 5.98  & 0.31  & 0.88 \\ 
13 & INSTEAD & Game & 0.41  & 0.74  & 0.54  & 0.80  & 7.62  & 4.79  & 0.44  & 0.94 \\ 
14 & ChaoChess & Game & 0.47  & 0.73  & 0.51  & 0.79  & 7.90  & 6.49  & 0.31  & 0.94 \\ 
15 & RailwaySP & Navig & 0.50  & 0.75  & 0.56  & 0.78  & 8.40  & 7.41  & 0.38  & 0.94 \\ 
16 & PocketMaps & Travel & 0.49  & 0.75  & 0.45  & 0.70  & 8.09  & 5.95  & 0.56  & 1.88 \\ 
17 & Barinsta & Internet & 0.50  & 0.74  & 0.52  & 0.77  & 5.50  & 3.65  & 0.44  & 1.00 \\ 
18 & FitNotif & Connect & 0.49  & 0.77  & 0.45  & 0.81  & 7.38  & 6.86  & 0.50  & 1.00 \\ 
19 & SkyTube & Media & 0.48  & 0.77  & 0.50  & 0.75  & 8.13  & 5.57  & 0.25  & 1.00 \\ 
20 & LibReader & Reading & 0.45  & 0.79  & 0.44  & 0.77  & 7.78  & 6.71  & 0.44  & 0.81 \\ 
\hline
\hline
\multicolumn{3}{c|}{\textbf{Average}} &  0.45  & \textbf{0.73}  & 0.49  & \textbf{0.79}  & 6.98  & \textbf{5.24}  & 0.54  & \textbf{1.33} \\ 
\hline
\end{tabular}
\end{table*}

\subsection{Experimental Design}
\label{sec_Experimental_Design}
To avoid potential inconsistency, we pre-install the 20 apps in the Google Nexus 6 on the emulator with Android 8.0 OS.
The experiment begins with a brief introduction to the task. 
We show a demo to participants in the experimental group about how to use our {\tool} with a new demo app (out of the 20 testing apps), and ask them to explore each app  separately. 
{\tool} will start to plan the test path for participants when they stay on the same page for 5 seconds without any operations.

The participants in the two groups need to test the 20 given mobile apps.
They are required to fully explore all the interfaces of each app and find as many bugs as possible.
Each participant has up to 10 minutes to test a mobile app which is far more than the typical app session (71.56 seconds)~\cite{bohmer2011falling}.
Each of them conducts the testing individually without any discussion with each other.
During their testing, all their screen interactions are recorded, based on which we derive their testing performance.

\subsection{Evaluation Metrics}
\label{sec_Coverage_Metrics}

Following previous studies~\cite{chen2019storydroid,chen2020improving}, the activity  and state coverage~\cite{su2017guided,li2017droidbot,cai2020fastbot} are common evaluation metrics in the Android GUI testing. So we use the following metrics to evaluate the effectiveness of {\tool}.
\begin{itemize}
\item State coverage: (number of discovered states) / (number of all possible states)
\item Activity coverage: (number of discovered activities) / (number of all activities)
\item Testing Time: average time spent per app (Each tester explore the entire app to the best of their ability)
\item Bug number: number of discovered bugs
\end{itemize}

\subsection{Results and Analysis}
\label{sec_Performance_Metrics}

We present the {\tool}'s average state and activity coverage, the average testing time, and the average detected bugs across the two groups, as shown in Table~\ref{tab:User study}.

\subsubsection{\textbf{Higher State and Activity Coverage}}
\label{sec_coverage}
As shown in Table~\ref{tab:User study}, the state coverage of the experimental group is 0.73, which is about 62\% ((0.73-0.45)/0.45) higher than that of the control group. 
And the activity coverage of the experimental group is 0.79, which is about 61\% ((0.79-0.49)/0.49) higher than that of the control group. 
The results of Mann-Whitney U Test shows there is a significant difference (p-value \textless 0.01, more detailed information of experiment can be seen in our website\textsuperscript{\ref{github}}.) between these two groups in both metrics.
It indicates that our {\tool} can guide the testers in exploring more states and activities possibly through reducing the duplicate explorations as well as the hesitation time in choosing which to explore next.
We also find that {\tool} can help testers spot some activities which are hard to find without the help of the tool.
We analyze those activities and summarize them into two categories.

\begin{figure*}[htb]
\centering
\vspace{0.1in}
\includegraphics[width=17.0cm]{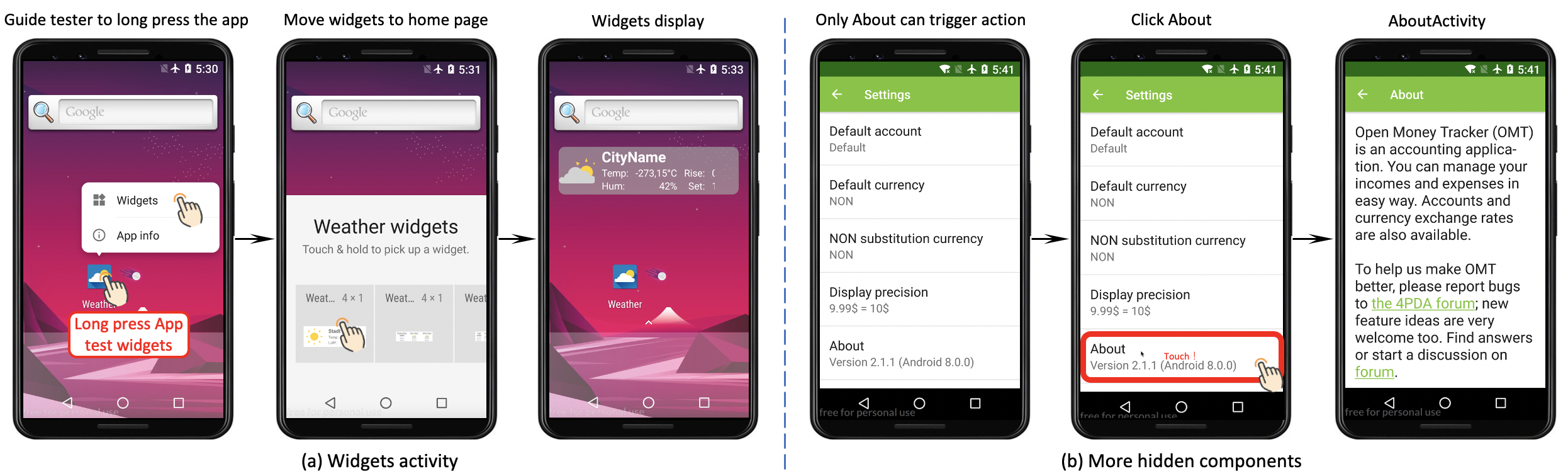}
\vspace{-0.1in}
\caption{Two difficult situations for manual testing}
\label{fig:result2example}
\vspace{-0.1in}
\end{figure*}

First, there are certain activities that require specific actions to trigger. 
For example, only \textit{long-pressing the app icon} can trigger the \textit{AppWidgetActivity}, which contains a pop-up showing the setting of the weather bar in Fig~\ref{fig:result2example} (a).
Second, some clickable components are inconspicuous due to the poor app GUI design.
Fig~\ref{fig:result2example} (b) shows another example, in which ``About'' is a button that can be clicked to show detailed information about the app version.
However, all other \textit{TextView} in the list cannot be clicked.
That is why all testers without using our tool missed it during manual testing.
These examples indicate that without explicit guidance as our {\tool}, human testers are very likely to miss those activities, resulting in the incompleteness of the app testing.

\subsubsection{\textbf{More Detected Bugs}}
\label{sec_results_RQ2_3}
With our {\tool}, testers can find an average of 1.33 bugs for each app, resulting in 30 unique bugs for the 20 experimental apps.
Due to the low activity coverage in the control group, some bugs are not discovered in this group, i.e., only an average of  0.54 bugs are reported for each app.
For example, as showed in Fig~\ref{fig:result2example} (a), testers did not notice the \textit{AppWidgetActivity}, thus could not explore the pop-up window which displays the setting of the weather bar. 
Therefore no one in this group finds the bug in the pop-up window.
The results of Mann-Whitney U Test shows there are significant difference (p-value \textless 0.01) between these two groups for the detected bugs.
Furthermore, these bugs are user interface bugs, which are difficult to be detected with current automatic testing techniques, further indicating the practical value of our proposed {\tool}.

In addition, for the control group, testers' attention and their reporting rate of bugs decreased after about 3 minutes of exploration during the process of the experiment, i.e., the duplication of testers' operation increased and the number of found bugs decreased.

\subsubsection{\textbf{Less Time Cost}}
\label{sec_results_RQ2_2}
It takes just 5.24 minutes for testers with our {\tool} to finish exploring an app by covering all pages while 6.98 minutes in the control group.
The results of Mann-Whitney U Test shows there are significant difference (p-value \textless 0.01) between these two groups for the testing time.
In fact, the average time of the control group is underestimated, because 14 participants fail to complete any apps even with 10 minutes, which means that they may need more time in exploring the apps.
 
We watch the video recording of the app testing in the control group for further discovering the reasons for the higher time cost.
Without the {\tool}, we find that testers repeated 0.69 (standard deviation $\sigma = 20\%$) of their own actions. In contrast, none of the testers using the {\tool} generated duplicate actions. 
This shows that the guide can effectively prevent testers from interacting with previously explored pages and components, and then guide them to explore deeper pages. 
It seems that the testers are stuck in some activities or activity loops without knowing how to trigger new activities. 
After the initial exploration of the app, some testers forget where they have explored, resulting in repeated testing of some activities.
This observation further confirms the importance of testing guidance tools as our {\tool}.


\section{Discussion}
\label{sec_discussion}
In summary, we find that the combination of our {\tool} can effectively guide testers to significantly increase their state and activity coverage. Consistent with Information Foraging Theory~\cite{pirolli1999information,lawrance2010programmers}, it suggests that providing visual navigation cues could help guide tester's attention and thus improve information access. 

\subsection{Testers' Experience With {\tool}}
According to the testers' feedback in Section \ref{sec_Usefulness}, all of them confirm the usefulness of our {\tool} in assisting their manual Android app testing.
They all appreciate that the hint moves of our {\tool} can help guide them in exploring the inconspicuous activities of the application, increasing the hit rate of potential bugs.
For example, ``{\tool} is very helpful for us to discover new pages and functions. It effectively avoids repeated operations.''(P1, P3, P8).
Participants express they like our interaction design such as ``Great, {\tool} uses the float window for guidance is a good idea. I like it!''(P2, P7, P13, P15, P6); ``The guidance is much like the tutorial when the game software is first used. It can help us to understand a new application.''(P4).
Participants express that our {\tool} can save their testing time such as``Nice! The {\tool} saves our testing time.''(P11, P16, P10).

The participants also mention the drawback and potential improvement of our {\tool}.
First, besides the hint moves highlighted along with their testing, they also hope to see the overall $STG_{action}$ before the application testing and also their real-time position in $STG_{action}$ during the Android app testing (P5, P9).
Second, more fine-grained transition states are needed as current ones can still not cover all functionality combinations or corner cases (P2,P3).

\subsection{Generalization of {\tool} and Findings}

{\tool} is designed to assist the manual testing of Android apps with the extracted STG on Android apps, and have achieved satisfactory performance.
In addition to Android, there are also many other platforms such as iOS, web.
To conquer the market, developers tend to develop either one cross-platform app or separated native apps for each platform considering the performance benefit of native apps.
Although our {\tool} is designed specifically for Android, it can also be extended to other platforms.
Given an app from other platforms, we just need an exploration tool for getting the STG.
With that STG, {\tool}'s path planning algorithm based on dynamic programming can be reused to facilitate the optimized exploration steps by avoiding repeated testing of UI page.

Therefore, we would expect that the idea of {\tool} can be applied to apps in other platforms.
Of course, the app usage pattern and the types of edits to predict are very likely to differ from one platform to another. For mobile platform like iOS, our empirical study and method may be easily adapted to it with some engineering effort.
For platforms using different devices like desktop, the differences between these platforms with Android can be considerably big.
In such cases, a detailed empirical study of the app usage pattern is required to determine the extension.
In the future, we will try to extract STG of multi-platform apps to guide downstream testing tasks. 

The {\tool} will also obtain a large number of UI screenshots in the process of assisting the crowdworkers to test the app, including issues such as UI display issue and compatibility issues. We will further study these screenshots in the future and realize automatic issue detection and repair.

\subsection{Potential Applications to End-users}
In addition to assisting human testers in GUI testing during the software development, our {\tool} can also be applied to help end-users in their daily app usage.
Given the increasing complexity of mobile apps, fluent usage of mobile apps is a challenging task especially for the aged and disabled users.
For example, there may be too many clickable components in one GUI page for senior users to locate the functionality which they want.
They may stuck on one page with repetitive attempts but not working.
Even for normal users, they may not notice some features especially news ones in the app. 

Based on our approach, according to the users' interaction log, our {\tool} can detect the situation of stuck mentioned above.
With the future algorithm improvement by machine learning, our approach can smartly remind users the next moves to jump out of the dilemma and explore new features.
In addition, our tool is running on users' device without any round-trip to a server, leading to the privacy preserving as privacy-sensitive data stays on the device.

\subsection{Limitations}
\label{subsec_Limitations}
\subsubsection{Incomplete $STG_{action}$}
Although our hybrid approach can construct an $STG_{action}$ with 74\% activity coverage, it still misses some states.
As analyzed in Section~\ref{sec_results_RQ1}, different developers would have varied code writing styles for implementing states and transitions among them, and some of them are with poor coding conventions.
All these could influence the complete extraction of $STG_{action}$.
We will keep improving our approach for covering more corner cases in enriching $STG_{action}$.

\subsubsection{Partial interaction and UI types}
In term of actions triggering the state transition in {\tool}, we only consider the clicking events, and could not handle other actions such as scrolling, text filling, etc.
Besides, it is also hard to visualize a complicated combination of actions on the same page to the next state. For UI pages with more animations (highly visual and dynamic app UI interfaces), {\tool} is difficult to operate on animations. The current method used by {\tool} is to skip these animation components.
Although those action combinations or animations only account for a small portion of state transition and state, we will also take them into consideration in the future.
\section{Conclusion}
\label{sec_conclusion}
As the last line of defence, manual testing is crucial to improve application quality.
Despite its importance, manual testing is time-consuming, labor-extensive and highly dependent on testers' experience and capability.
Therefore, we propose a method called {\tool} to guide human testers in exploring more states during app testing.
We first construct $STG_{action}$ by both static analysis and dynamic exploration, and generate the planned path based on a dynamic programming algorithm.
On the app screen, we highlight the hint moves triggering the next unexplored state to users by a visualized floating window of that page.
The automated evaluation and user study demonstrate the accuracy and usefulness of {\tool} in improving testing efficiency, reducing testing time and saving testers' efforts.

In the future, we will work in two directions.
First, we will improve our approach in constructing enhanced $STG_{action}$ by detecting more states and fine-grained transitions among them.
According to the user feedback, we will also improve the interaction between our {\tool} and users, which can borrow the idea from the human-machine collaboration studies to better facilitate the human testers.
Second, we will not limit our {\tool} to app testing, and plan to explore its potential usage into other areas like reminding users about the new functionalities of apps.

\begin{acks}
This work is supported by the National Key Research and Development Program of China under grant No.2018YFB1403400, National Natural Science Foundation
of China under Grant No. 62072442, No. 62002348, and Youth Innovation Promotion Association Chinese Academy of Sciences.
\end{acks}


\bibliographystyle{ACM-Reference-Format}

\bibliography{reference}

\end{document}